# A Strategy to Identify Components Using Clustering Approach For Component Reusability

N Md Jubair Basha[1] and Dr Chandra Mohan[2]

[1]Assistant Professor, IT Department, Muffakham Jah College of Engineering & Technology Hyderabad, India.
jubairbasha@mjcollege.ac.in

[2] Associate Professor, CSE Department, JNT Univerity College of Engineering, Hyderabad, India.
c_miryala@yahoo.com

## ABSTRACT

*Component Based Software Engineering (CBSE) has played a very important role for building larger software systems The current practices of software industry demands development of a software within time and budget which is highly productive. It is necessary to achieve how much effectively the software component is reusable. To achieve this, the component identification is mandatory. The traditional approaches are presented in the literature. However effective reuse is still a challenging issue. In this paper, a strategy has been proposed for the identification of a business component using clustering methodology. This approach will be useful in identifying the reusable components for different domains. The proposed approach has identified the reconfigured component using the CBO measure to reduce the coupling between the objects. By considering this proposed strategy, the productivity can be increased in the organization.*

## 1. INTRODUCTION

Component Based Software Engineering (CBSE) has played a very important role for building larger software systems. Reduction of Cost [1] and shorter development i.e. within time gives a good prospect for increasing the productivity in the organization. Components are connected by assembling, adapting and wiring into a complete application. Although there is no IEEE/ISO standard definition that we know of, one of the leading exponents in this area, Szyperski [2], defines a software component as follows:

*"A software component is a unit of composition with contractually specified interfaces and explicit context dependencies only. A software component can be deployed independently and is subject to composition by third parties".*

Effective software reuse helps in the development of quality product within time and budget. This also helps in reducing the high effort needed for testing and maintenance of the software products.

Several approaches are proposed in the literature which provides only subset of operation requirements of effective software reuse. Though some of the approaches identify the components[21] as an artifact, but they may won't identify business components. In this paper, a strategy to identify the components using clustering is proposed for the effective software reuse. The remaining part of this paper is organized as follows: section-2 presents the advantages of reusing software systems, section-3 describes about the domain engineering with its process and approaches for the domain engineering, section-4 describes the proposed strategy for the identification of components using clustering methodology which maintains low coupling and high cohesion. This calculates the component reconfiguration using CBO measure and section -5 concludes the paper.

## 2. SOFTWARE REUSE

Software Reuse is the use of available software or to build new software from software knowledge. Reusable assets can be either reusable software or software knowledge. Reusability is a property of a software asset that indicates it's probability of reuse [3]. Software Reuse means the process that use "designed software for reuse" again and again [4]. By reusing software, we can manage complexity of software development, increase product quality and makes faster production in the organization.

Recently, design reuse has become popular with (object-oriented) class libraries, application frameworks, design patterns and along with the source code [5]. Jianli et al. proposed two complementary methods for reusing existing components. Among them one allows component evolution itself, which is achieved with binary class level inheritance across component modules. The other is by defined semantic entity so that they can be assembled at compile time or bind at runtime. Although component containment still is the main reuse model that leads to contribute the software product lines development [6]. Regarding the components much information has to be collected, maintained and processed for the retrieval of the components. Maurizio has proposed a methodology to automatically build a software catalogue tools for archiving and retrieval of information are presented [7]. Software Reuse can be broadly divided into two categories viz. Product reuse and Process reuse. The product reuse includes the reuse of a software component and by producing a new component as a result of module integration and construction. The process reuse represents the reuse of legacy component from repository. These components may be either directly reused or may need a minor modification. The modified software component can be archived by versioning these components. The components may be classified and selected depending on the required domain [8].

The construction of a components is fundamental to their use. Reuse does not come as a side effect. Specification, construction and testing must all be done for reuse. This makes a component more expensive (up to 10 times) to develop a new software.

Several different criteria for a good component have been suggested. These criteria can be summarized in the following:

- The component should represent an abstraction. It should have high cohesion and offer only the operations needed to make it useful in an efficient manner. It must also have well defined interface, both syntactically and semantically. If two operations in two different components have the same name, they should act in a similar manner. But their style should be similar to facilitate **understanding.**

- The component must be **independent** of surrounding entities; it should be loosly connected and thus have low coupling to other units. An object-oriented philosophy leads to independence.

- The component should be **general** abstraction which is useful in several applications without having to go unnecessary changes.

Understandability must be internal as well as external. Since good components will have a long life, they will be maintained for a long time.

The component system includes the selecting, classifying and managing the components included in the repository and also the development of new components. The component repository should be spread throughout the development organization and that the components are accessible. The component repository should preferably be shared between several different products. It means that the component system should serve several projects. Whenever the new projects to be taken up, then the relevant components shall be needed for the development process. The project proposals should be reviewed by a group consisting of experienced designers and also someone from component department forming a software component committee. They should judge whether the proposed components are needs to be developed or not. If it is decided to construct the component, it is forwarded to component construction with a deadline. When ready, it is added to the component repository which then takes a new version state as showed in the Figure 1. As component is being used, the software component group should analyze it's value. Which component is used most? Which are not used at all? How much you gain from the components? This analysis helps to develop the component system.

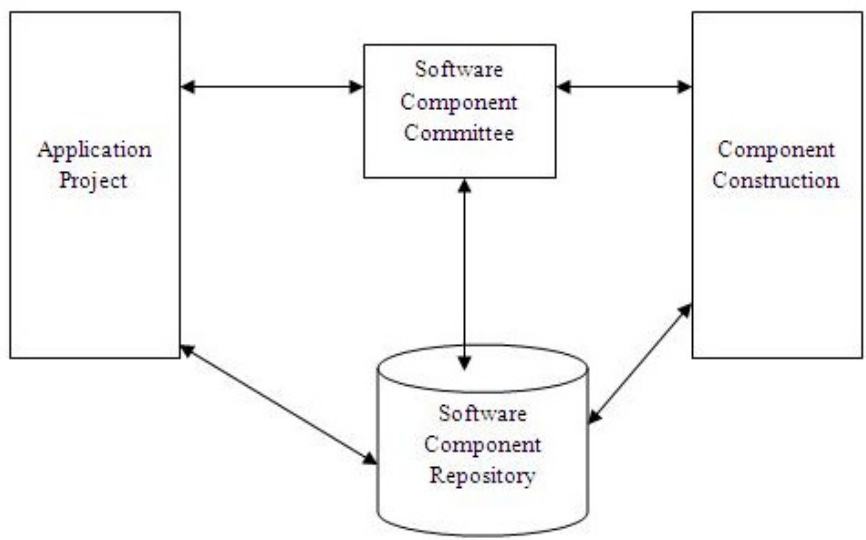

Figure: 1. An Organization for Component Management

## 3. DOMAIN ENGINEERING

Software Reuse can be improved by identifying objects and operations for a class of similar systems, i.e. for a particular domain. In the context of software engineering domains are application areas [9].

There are various definitions of what a *domain* is. Czarnecki's defines [10]:" an area of knowledge scoped to maximize the satisfaction of the requirements of stakeholders, which includes concepts and terminology understood by practitioners in the area and the knowledge of how to build (part of) systems in the area".

Domain Engineering is a process in which the reusable component is developed and organized and in which the architecture meeting requirements of the domain is designed [11].

Domain Engineering can be defined by identifying the candidate domains and performing domain analysis and domain implementation which includes both application engineering and component engineering. Domain Analysis is a continuing process of creating and maintaining the reuse infrastructure in a certain domain. The main objective of domain analysis is to make the whole information readily available. The relevant components (if available) has to be extracted from the repository rather than building the new components from the scratch for a particular domain.

Domain Analysis mainly focuses on reusability of analysis and design, but not code.This can be achieved by building common architectures, generic models or specialized languages that additionally improve the software development process in the specific problem area of the domain. A vertical domain is a specific class of systems. A horizontal domain contains general software parts being used across multiple vertical domains. Mathematical functions libraries container classes and UNIX tools are the examples of horizontal reuse. The purpose of domain engineering is to identify objects and operations of a class in a particular problem domain [9].

In the process of domain analysis, each component identified can be categorized as follows.

- General-purpose components : These components can be used in various applications of different domains (horizontal reuse).
- Domain-specific components :They are more specific and can be used in various applications of one domain (vertical reuse).
- Product-specific components : They are very specific and custom-built for a certain application, they are not reusable or only useful to a small extent.

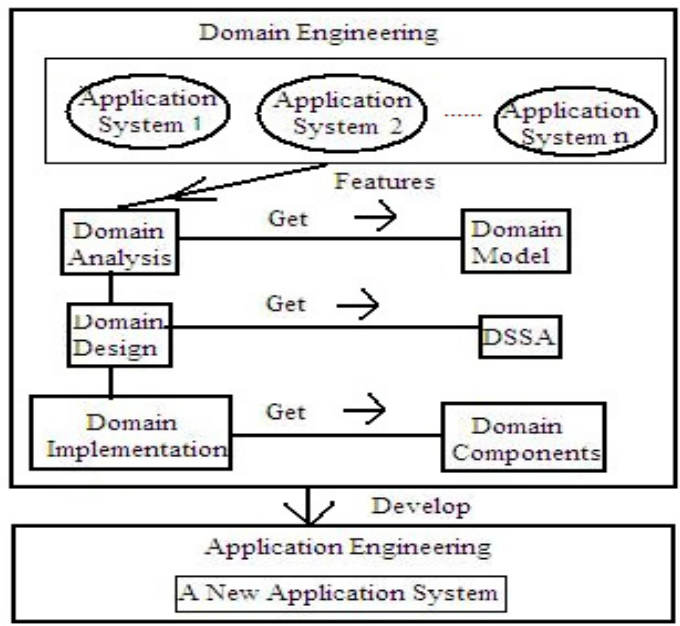

Figure:2. Domain Engineering Process

Domain engineering process [18] is depicted in figure 2. DE consists of three main stages i.e. domain analysis, domain design and domain implementation. For Domain Analysis support, DARE-COTS tool is presented [3]. Initially, in a particular domain it is mandatory to get the universal and variable characteristics of group systems. By abstracting the characteristics, domain analysis model can be generated.  Based on this model the domain specific software architecture can be designed and then reusable components will be generated and organized.

### 3.1. Approaches For Domain Engineering

There are several known domain engineering tools. Each of these tools specifies a subset of operation requirements

- Domain Analysis and Reuse Environment (DARE) is a tool developed in 1998 to support capturing information from experts, documents and code. Captured domain information is stored in a database that typically contains a generic architecture for the domain and domain-specific reusable components. DARE provides a library search facility with a windowed interface to retrieve the stored domain information [12].
- Family-Oriented Abstraction, Specification and Translation (FAST) is a system family generating method based on an application modelling language(AML) and was guiding developers to create the tools needed to generate software product line using domain engineering phase and application engineering phase.
- Feature Oriented Reuse Method (FORM) as an extension to the Feature Oriented Domain Analysis (FODA), a systematic method of capturing and analyzing commonalities and differences of applications in a domain (features). By using the results to develop domain architectures and components and modelling to discover and understand and capture commonalities' and variability's of a product line [13].
- Kobra (KomponentenbasierteAnwendungsentwicklung) is used for component-based development [3]. Kobra method consists of product line development, component based software development and frameworks to provide systematic approach to developing high quality component based application frameworks [14]. Kobra is "technology independent" in the sense that it can be used with all the three major component implementation technologies CORBA, Java Beans and COM.
- Product Line UML-Based Software Engineering (PLUS) is a model-driven evolutionary development approach for software product lines. Apart from the analyzing and modelling a single system, it provides a set of concepts and techniques to explicitly model the commonality and variability in a software product line. With these techniques, object oriented requirements, analysis and design models of software product lines are developed using UML 2.0[15].
- Component Oriented Reverse Engineering (CORE) is a systematic and concrete model used to identify and develop reusable software components by using the reverse engineering techniques. This is used to extract architectural information and services from legacy system and later on convert the services into components [16].

## 3. A PROPOSED STRATEGY FOR COMPONENT IDENTIFICATION

Researchers from 1990, Component Identification has been considered as a sole problem and wide issue. Considering from software reuse cost optimization, researchers focused to cluster the business models according to the "high cohesion and low coupling" strategy and combined each cluster into a component [17].

The fundamental methodology is as follows:

1. Calculate the strength of semantics dependencies between two business elements.

2.  Transforms business models into the form of weighted directional graph, in which business elements are categorized as nodes and semantics dependency strength are the weight of the edges between two nodes.
3.  Cluster the graph using clustering or matrix analysis techniques.

The above explained methodology [18] is known as Clustering Analysis in Mathematical statistics for precise classification. It aggregates those elements with high cohesion together to form specific patterns, which is widely used in the field of data mining and pattern recognition. The above methodology can be imported into Component Identification and expected to obtain components with high cohesion and low coupling to reduce composition cost. Depending on different strategies of calculating dependency strength (DS) between nodes, clustering analysis may produce different results. The fundamental methodology is presented as follows:

1.  Denote $n$ elements that need to be classified as set N, and initially each element in N form a cluster.
2.  Specify the principles for calculating Dependency Strength (DS) i.e. similarity between arbitrary two nodes, and denote DS between $N_i$ and $N_j$ as $F_{ij}$,
3.  Calculate DS between arbitrary two nodes in N and obtain the DS metric D of $n$ elements.
4.  Choose a sound "Minimum DS" $F_{min}$ as the judgment principle for merge two elements into one cluster.
5.  According to $F_{ij}$ in D, execute the following clustering process.
    5.1. ( Value value) if $F_{ij} >= F_{min}$, then set $N_i$ and $N_j$ into one cluster.
    5.2. (Transitivity) if $N_i$ and $N_j$, $N_i$ and $N_k$ belong to the same cluster respectively. Then merge $N_i$ $N_j$, $N_k$ into one cluster.
6.  Map elements in each cluster together into a business component.

With this methodology, it is possible to achieve the clustering for the identification of business components. By considering low coupling and high cohesion, component reusability can be easily realized.

The proposed strategy, has realized using an HR Portal application. This application includes three components i.e. Web-Tier(WBR), Business-Tier(BR), Data Access Object(DAO). These three components consist of 13 classes. Web tier component contains 5 classes, business tier contains 3 classes and DAO contains 5 classes.

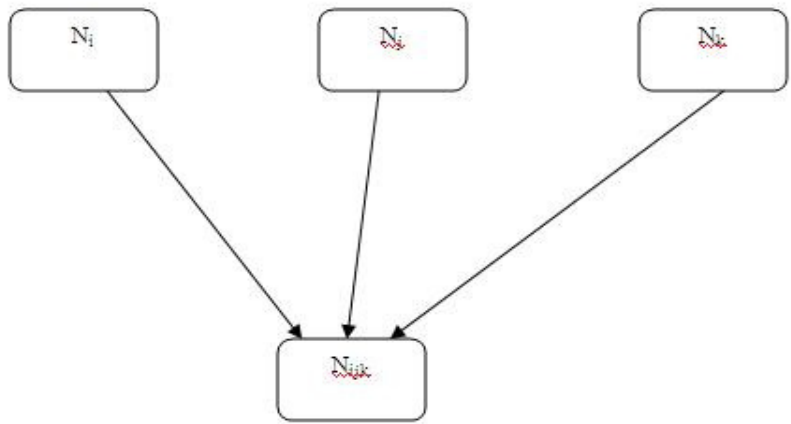

Figure: 3 Merging of nodes to a single cluster

If the object contains one method, then there exists one initial and one final node. If the object contains two methods then there exist four possible nodes[20]. The nodes depend on the sequential execution of the methods. Based on the execution logic of a method the object might change to another or might remain with the same node or with different nodes.

The following table depicts the no. of nodes transitions for different methods: m1(), m1( m2()), m2(), m2(m1()).

At any particular point of time, if the designer wants to know about which part of the system is not effectively reused then a lookup is to be performed on the component management relation. A Central repository maintains a table for managing a component reuse. This table contains three fields. One field specifies the name of the component and the second field contains the count specifies the number of times the component was reused by several systems. The last field contains the mapping of the nodes with the related components.

Table 1. Component Management Relation

| Component | Count of Reuse | Node |
|---|---|---|
| Webtier(WBR) | 24 | $N_i$ |
| Businesstier(BR) | 10 | $N_j$ |
| DAO | 36 | $N_k$ |

The above strategy is considered for the component identification. By following this strategy, an algorithm has to be developed so that the results for the different systems can be made for the component identification. This idea is to identify the reusable components which are highly cohesive.

With this, component identification techniques, the reusable component can be measured by considering the metrics for the components as proposed in [19]. The components which are highly cohesive needs to be reconfigured. Coupling Between Object Measure (CBOM) is used to identify the highly cohesive components.

Coupling between object measure (CBOM) for a component is defined as the number of invocations by the specified component. Those components whose CBOM is high or those component(s) of the system whose CBOM is greater than certain scalar value are the components which needs to be reconfigured at the earliest.

Hence a Reconfigurable Component(Cr) can be identified as follows

Cr = Max { CBOM(Ci) }, where Ci=C1, C2,... Cn are components of the system (or)

$Cr_j$ = { CBOM(Ci) > P}, where P is a scalar value whose value differs from one domain to the other.

Based on the invocations specified in the HR Portal application[9], CBOM is evaluated to identify one reconfigurable component.

As there are three components of the HR Portal application viz., Web-tier, Business-tier and DAO, the component with highest CBOM is the candidate for reconfigurable component.

In order to understand, the Coupling Between Objects Measure of Web tier component as well

Coupling Between Objects Measure of of Business tier and Coupling Between Objects Measure of DAO are considered for identification of Reconfigurable Component(Cr).

Hence,

Cr = Max { CBOM(WBR), CBOM(BR), CBOM(DAO) }

Cr = Max { 180, 95,224}= 224

As Coupling between Object Measure of DAO component is maximum, it is needed to reconfigured.

Hence the DAO component has to further reconfigured to reduce the coupling. This is achieved by dividing the DAO component into two sub DAO components viz., $DAO_1$ & $DAO_2$. The division of the DAO component makes the DAO component less cohesive. This is realized later by evaluating the CBO measure of $DAO_1$ & $DAO_2$.

## 5.CONCLUSIONS

As there is a need to identify the business reusable components, the proposed strategy has given an idea for the component identification by using clustering method. In order to identify the components, it is necessary to maintain the high cohesion and low coupling. Whenever the components are loosely coupled it is easy to extract. This criteria will increases the reusability of the component. As a future scenario, it is needed to implement this strategy  to realize on any domain with some results. By identifying the reconfigured component, helps    to reduce the coupling between objects using  CBO measure.  By considering this strategy , the productivity can be easily increased in the organization.

## ACKNOWLEDGEMENTS

The work was partly supported by the R & D Cell of Muffakham Jah College of Engineering & Technology, Hyderabad, India. The authors would like to thank to all the people from Industry and Academia for their active support.

**Authors**

N Md Jubair Basha received his B.Tech. (IT) and M.Tech (IT) from JNTUH, Hyderabad. He is presently working as Assistant Professor in Department of Information Technology, Muffakham Jah College of Engineering and Technology, Hyderabad, India. His research interest includes Software Reusability, Component Based Software Development, Data Mining and Cryptography. He has published many research papers in various National/ International Conferences and Journals. He is an active member of IEEE and CSI.

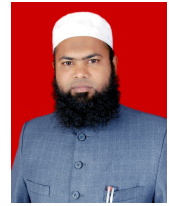

M. Chandra Mohan received B.E. (EEE) degree from Osmania University in 1994. He worked as Assistant Engineer in AP State Electricity Board (APSEB) for 7 years (1994-2001). He completed his    M.Tech. (CS&E) from Osmania University in 2000. He is working in JNT University Hyderabad since 2001. Presently he is working as an Associate Professor in Dept of CS&E in JNTUH College of Engineering Hyderabad, JNT University Hyderabad. He is the recipient of 3 Gold Medals from Osmania University at the graduate level by securing University first rank. He completed his Ph.D in 2010 from JNTU Hyderabad in Computer Science & Engineering. He has published 12 research publications in various National and International Journals and conferences

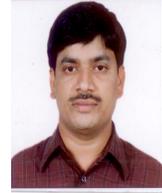